\documentclass[bibauthoryear]{aa}  
\usepackage{graphicx}
\usepackage{sidecap}

\usepackage{txfonts}
\makeatletter
\renewcommand\@biblabel[1]{}
\makeatother
\begin{document}
  \title {The internal structure of neutron stars and white dwarfs, and the 
Jacobi virial equation. II}
\author{ A. Claret\inst{1} \and M. Hempel \inst{2}}
 \institute{Instituto de Astrof\'{\i}sica de Andaluc\'{\i}a, CSIC, 
Apartado 3004, 18080 Granada, Spain  \\ 
\email{claret.iaa.es}
\and
Department of Physics, University of Basel, Klingelbergstr. 82, 
4056 Basel,
Switzerland }     
\date{Received 15 October 2012 / Accepted 30 January 2013}
\abstract
   { The Jacobi virial equation is a very powerful tool in exploring several aspects of 
the stellar internal structure and  evolution. In a previous paper we have shown that 
the function $\left[\alpha\beta\right]_\mathrm{GR}/\Lambda^{0.9}(R)$ is constant 
($\approx 0.4$) for pre main-sequence stars (PMS), white dwarfs (WD) and for some 
neutron star (NS) models, where $\alpha_{GR}$ and $\beta_{GR}$ are  the form-factors of 
the  gravitational potential energy and of the moment of inertia.}
 { To investigate  the structural evolution  of another type of celestial bodies, 
we extend these calculations to gaseous planets.  We  also analyse the cases for which  
this  function is not conserved during some stellar evolutionary phases.  
Concerning NS, we study the influence of the equation of state (EOS) on this 
function and refine the exponent of the auxiliary function $\Lambda(R)$.  
We also present  a macroscopic  criterion of stability for these stars.   }
   { Non-stop calculations from the pre main-sequence   to the white dwarf  cooling 
sequences were performed with the MESA code. The covered mass range was  0.1-1.7 
$M_\odot$.  We used the same code to compute models for  gaseous planets with masses 
between 0.1- 50  $M_J$.  Neutron star  models were computed using two codes. The 
first one is a modified version of the  NSCool/TOV subroutines. The second code is a plain 
TOV solver that allows one to use  seven previously described EOS. The relativistic  moment 
of inertia and gravitational  potential energy were computed  through  a fourth-order 
Runge-Kutta method.    }
   {By analysing  the internal structure of gaseous planets  we show that the 
function $\left[\alpha\beta\right]_\mathrm{GR}/\Lambda^{0.8}(R) \equiv \Gamma(M, \text{EOS})$ is 
conserved  for all models during the whole  planetary evolution and is independent of the 
planet mass. For the pre main-sequence   to the white dwarf  cooling 
sequences, we have found a connection between the strong variations of  $\Gamma(M, \text{EOS})$ 
during the intermediary evolutionary phases and the specific nuclear power. A threshold 
for the specific nuclear power was  found. Below this limit this function is 
invariant ($\approx 0.4$) for  these models, i. e., at the initial and final stages 
(PMS and WD). 
For NS, we showed that the function $\Gamma(M, \text{EOS})$  is also invariant 
($\approx 0.4$) and  is independent of the EOS and of the stellar mass. Therefore, we 
confirm that regardless of the final products of the stellar evolution, NS or  WD, they  
recover the initial value of  $\Gamma(M, \text{EOS})  \approx 0.4$ acquired at the PMS. Finally, 
we have introduced  a  macroscopic stability criterion for neutron star models based 
on the properties  of the relativistic product $\left[\alpha\beta\right]_\mathrm{GR}$. 
  }
   {}

   \titlerunning {The internal structure of neutron stars and white dwarfs  }
   \maketitle
%

\section{Introduction}

The vectorial equations often used in  physics are sometimes too complicated to  be solved 
in a closed form.  One of the advantages of the virial theorem is that it reduces the 
complexity of these equations,  but this frequently comes at the expense of losing information. The 
balance is, however, positive because some  complicated problems can be treated and 
important information can be derived. 
The moment of inertia and the gravitational  potential energy are fundamental elements  
 that are handled by the Jacobi virial equation. This equation is a very useful tool for 
exploring several aspects of the stellar internal structure and  evolution. Given the nature 
of this equation, a mathematical relation between the moment of inertia and  
the gravitational potential energy reduces its complexity, which facilitates managing it. 
Ferronsky et al. (1978) argued that this relationship  could be expressed by the constancy 
of the product $\alpha\beta$ in the Newtonian approximation, where $\alpha$ and $\beta$ 
are the form-factors of the gravitational potential energy and of the moment of inertia,  
respectively. However, the laws of mass distribution used by these authors  to evaluate 
the product $\left[\alpha\beta\right]_\mathrm{Newt}$  covered only a small branch of the 
Herzsprung-Russel (HR) diagram.  To check the constancy 
of $\left[\alpha\beta\right]_\mathrm{Newt}$ by using  realistic stellar models, Claret \& 
Gim\'enez (1989)  provided values of  $\alpha$ and $\beta$ for several grids of stellar 
evolutionary models. The availability of these form-factors allowed deeper investigations 
into the Jacobi dynamics of stellar evolution (Quiroga \& Claret 1992ab; 
Quiroga \& Mello 1992, Quiroga \& Cerqueira 1992). Unfortunately, due to limitations in 
the  opacity tables available at that time, these stellar models only covered the pre 
main-sequence, the main sequence and only the first stages of the red giant branch. 
Therefore, many important aspects of the later stellar evolution could not be 
investigated in the mentioned papers.  Another considerable limitation of these 
papers concerns mass loss, which is not considered. As we will see in the next 
sections, this point is very important. For example, a model with an  initial mass of 
1.7 $M_\odot$ at the pre main-sequence  reaches the white dwarf stage with  only 
0.6 $M_\odot$. Modern stellar models available in the literature  improved the 
situation:  Claret (2006, 2007) published grids covering a wide range of chemical 
compositions (Z= 0.001-0.10), masses (0.8-125 $M_\odot$)  and also offered detailed  tracking  of the late phases of stellar evolution.

In a recent paper Claret (2012)   showed that the product of the form-factors $\alpha$ 
and $\beta$ -- neither the Newtonian nor the relativistic one -- is not  conserved 
 during the  core hydrogen-burning phase, helium-burning, thermally pulsating  
asymptotic giant branch, or 'blue loops' for models evolving from the pre main-sequence 
 to the white dwarfs stages.  In the same paper, it was shown that for very compact 
objects (neutron stars  (NS) - for example) the effects of  general relativity must be 
taken into account and the product $\alpha\beta$ computed  using the Newtonian
approximation  must be superseded by the relativistic product 
$\left[\alpha\beta\right]_\mathrm{GR}$.

In  our  formulation the conserved quantity at the beginning and at the end of the
stellar evolution   is  the function 
$\left[\alpha\beta\right]_\mathrm{GR}/\Lambda^{0.8}(R) \equiv \Gamma(M, \text{EOS}) 
\approx$ 0.4. 
 The   auxiliary function 
$\Lambda(R)$ is given  by $\left[1 - \frac{2 G M(R)}{R c^2}\right]^{-1}$ 
and is evaluated at the surface of each model, $c$ is the speed of light, $R$ is
the radius, $M(R)$ is the  gravitational mass, and $G$ is the constant of
gravitation. Claret (2012) has also shown  that, regardless of the final products of stellar 
evolution (white dwarfs  (WD) or NS),  they recovered the fossil value of 
  $\left[\alpha\beta\right]_\mathrm{GR}/\Lambda^{0.9}(R)$ acquired at 
the pre-main sequence (PMS) phase.  The refinement of the exponent in the quoted function 
using an extended set of more elaborate EOS is given in Sect. 5. 

In the present paper we explore some aspects connected to these 
investigations. First, we extend the calculations of the moment of inertia and  the 
gravitational potential energy to gaseous planets and  show that the function  
$\left[\alpha\beta\right]_\mathrm{GR}/\Lambda^{0.8}(R)$ (hereafter $\Gamma(M, \text{EOS})$) is 
conserved during the planetary evolution, regardless of the initial mass. Second, 
we investigate why  this function is not conserved during some evolutionary 
phases  when we consider a complete evolution  from the PMS  to the WD  cooling 
sequences  (hereafter PMS-WD models). Third, we confirm that $\Gamma(M, \text{EOS})$ is 
conserved for NS and, in addition, is independent of the equation of state and
of the stellar mass.  We also derive, heuristically, an alternative criterion of 
stability based on the properties of the gravitational potential energy and of the 
moment of inertia.

\section {Neutron star, pre main-sequence, white  dwarf, and gaseous
planets models}

\subsection {Planets}

The models of gaseous planets were computed using the MESA code (Paxton et al.
2011,  version 4298) for a chemical composition of X=0.73 and Z=0.01. The adopted
mixing-length parameter $\alpha_{MLT}$ is 1.5. We computed planet models from 0.1 
up to 50 $M_J$. All models were followed from the gravitational contraction  to an 
approximate age of 20 Gyr. 

\subsection {Non-stop calculations: from pre main-sequence to white dwarf
models}

We  used the same version of the MESA code to compute a non-stop evolution from the PMS to 
WD cooling sequences.  The adopted chemical composition and the  mixing-length parameter 
$\alpha_{MLT}$ are  the same as those adopted for the planet models. 

\subsection {Neutron  star models}

The neutron star models were computed using two different codes. The first  
is a modified version of the  NSCool/TOV subroutines (Page \& Reddy 2006; Page,
Geppert \& Weber 2006).  We  computed neutron star models  adopting the 
non-relativistic  EOS (equation of state) by Akmal, Pandharipande  \&  Ravenhall (1998) (APR).

Another set of subroutines  was used to explore the role of the EOS in the invariance 
of the function $\Gamma(M, \text{EOS})$.   We considered seven different EOS  that were 
calculated with the model from Hempel \& Schaffner-Bielich 2010, using the 
following  relativistic mean-field nucleon interactions:  $DD2$ (Typel et al. 2010), 
$FSUgold$ (Todd-Rutel \& Piekarewicz 2005), $NL3$ (Lalazissis,  K\"onig \&  Ring 1997), 
$SFHo$, and $SFHx$ (Steiner et al. 2012), $TM1$ (Sugahara \& Toki 1994),  and $TMA$ 
(Toki et al. 1995). Some subroutines were added to compute the apsidal-motion 
constants, the moment of inertia, and the gravitational  potential energy. 

As we have shown in Claret (2012),  the effects of  general relativity  on 
the moment  of inertia and gravitational potential energy are not important for PMS, 
main-sequence (MS) stars,   WD and, as we will see below,  for  planets. 
However,  for consistency, we used the relativistic formalism throughout.  The moment of 
inertia $I_{GR}$ and the  gravitational potential energy $\Omega_{GR}$ were computed using the
following equations (for the approximation of the moment of inertia, see  Ravenhall \&  
Pethick 1994 and for the gravitational potential energy, see Misner, Thorne \& Wheeler 1973 ):  

\begin{flalign}
&{J_{GR} = {\frac{8\pi}{3}}\int_{0}^{R} \Lambda(r)r^4\left[\rho (r) +
P(r)/c^2\right] dr }, \nonumber & \\
&I_{GR} \simeq  \frac{J_{GR}}{\left(1 + \frac{2GJ_{GR}}{R^3c^2}\right)} \equiv
{(\beta_{GR} R)^2}M & \\
&{\Omega_{GR} = -4\pi\int_{0}^{R} {r^2\rho\left[\Lambda^{1/2}(r) - 1\right]dr} } 
\equiv -{\alpha_{GR}} \frac{G M^2}{R}, & 
\end{flalign}

\noindent
where  $P(r)$ is the pressure, $\rho(r)$ the energy density and the auxiliary 
function $\Lambda(r)$ is given by  $\left[1 - \frac{2 G m(r)}{r c^2}\right]^{-1}$. 
Equations (1)  and (2) were integrated through a  fourth-order Runge-Kutta method.

  \begin{figure}
   \includegraphics[angle=-90,width=\columnwidth, totalheight=0.94\columnwidth]{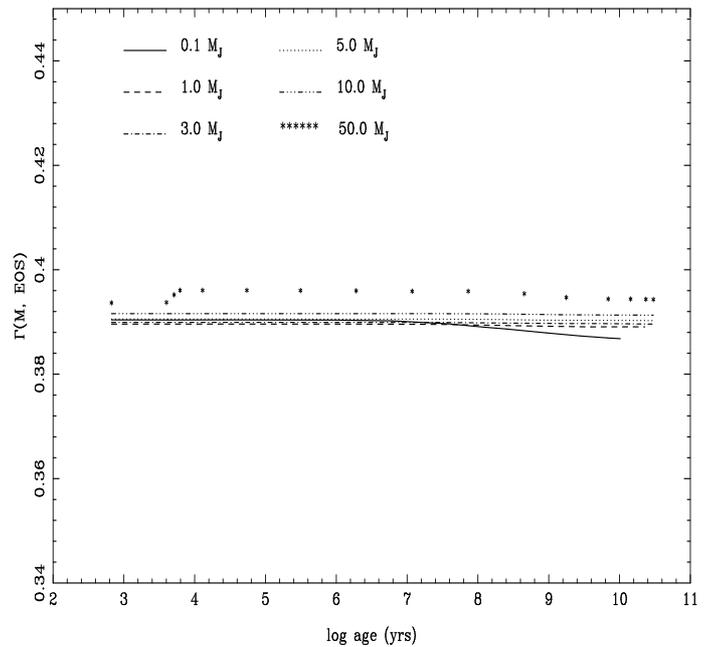}
   \caption{Time evolution of  $\Gamma(M, \text{EOS})$  
for some planets with masses between 0.1 and 50 M$_J$. }
   \end{figure}

 \begin{figure*}
   \includegraphics[angle=-90,width=\hsize]{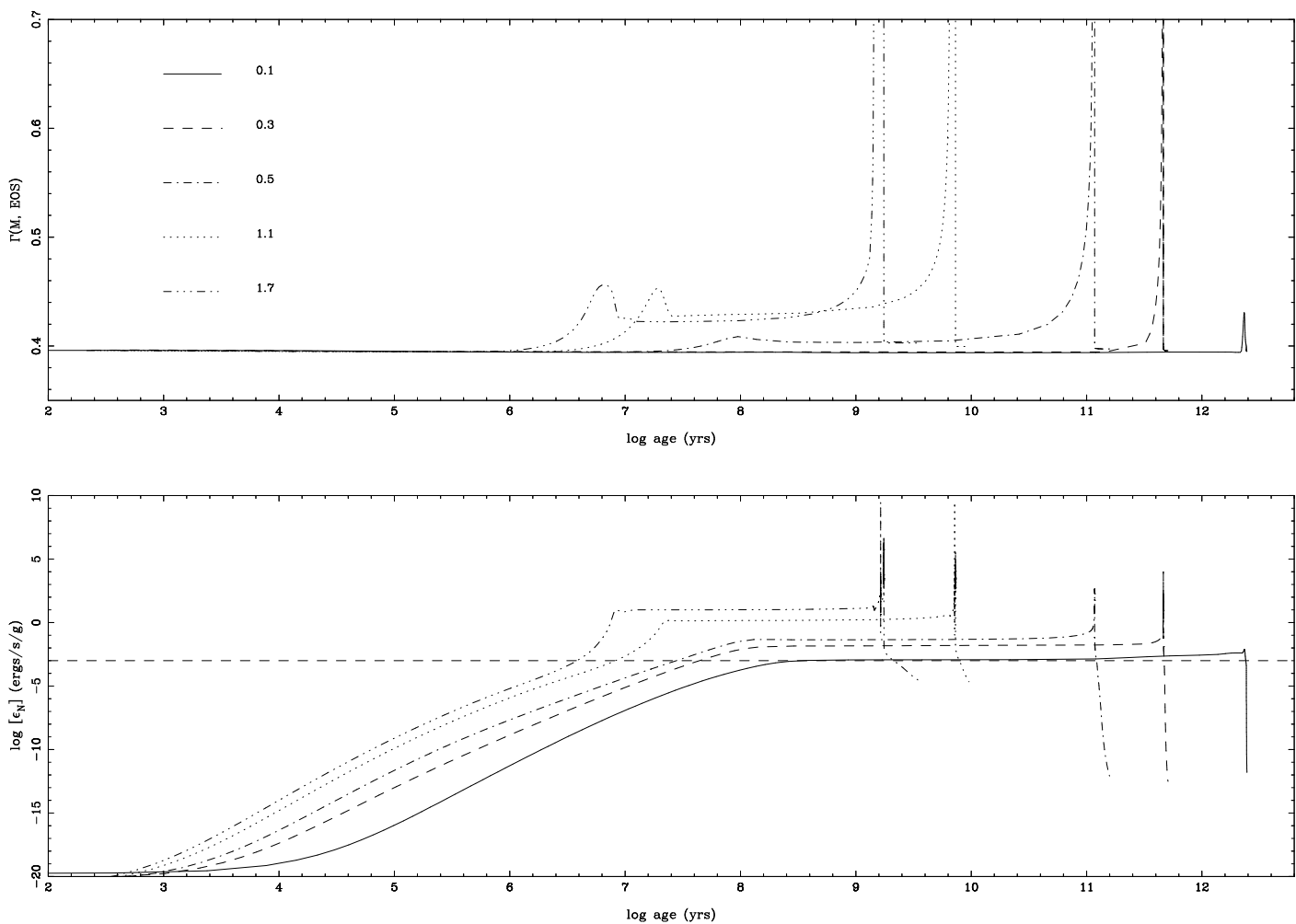}
   \caption{First panel: time evolution of  $\Gamma(M, \text{EOS})$   
for some stellar models evolving  from the PMS  to the WD stage. The mass
range is 0.1--1.7  $M_{\odot}$ (see symbols in the left upper corner). The adopted 
initial chemical composition  is X = 0.73, Z= 0.01 and the mixing-length parameter 
$\alpha_{MLT}$ = 1.5.  Second panel: specific nuclear power for the same models.}
   \end{figure*}

\section {Gaseous planets}

Figure 1 displays the calculations of $\Gamma(M, \text{EOS})$  for gaseous planets with masses 
ranging from 0.1 up to 50 $M_J$.   Although $\Lambda(R)$ tends to 1.0 for all gaseous 
planet models, for consistency,  we kept the definition of $\Gamma(M, \text{EOS})$. The 
refinement of the exponent in the above function is explained in Sect. 5.  The evolution 
of each model was followed from the gravitational contraction to an approximate age 
of 20  Gyr. 

An inspection of Fig. 1 shows that $\Gamma(M, \text{EOS})$ is constant for all evolutionary 
phases  ($\approx$ 0.4). This is the same value we obtained for PMS, WD and NS models 
(Claret 2012). We performed some tests by varying the initial chemical composition and 
$\alpha_{MLT}$  and the results are independent of the input physics. With these 
calculations we extended  the invariance of $\Gamma(M, \text{EOS})$ to another category of 
celestial bodies, the planets.  

As we show below, the invariance of the aforementioned function during all 
planetary evolutionary phases is connected with the specific nuclear power that, for 
the gaseous planets we analysed here, is below a threshold that is determined in the
next section. 

\section {Connection between the  specific nuclear power 
$\epsilon_N$ and $\Gamma(M, \text{EOS})$ for PMS  to the WD cooling sequences}

In a previous paper  (Claret 2012) we argued that the drastic changes in 
$\alpha\beta$ during the MS,  thermally pulsing asymptotic giant branch, and "blue loops" 
for PMS-WD models could be related to the presence/absence  of nuclear reactions,  with
chemical inhomogeneities, etc. To try to elucidate this point we  computed some 
PMS-WD models with masses varying from 0.1 up to 1.7 $M_{\odot}$. 
We  found that the specific nuclear power $\epsilon_N$ is directly
correlated with the variation of   $\Gamma(M, \text{EOS})$, as shown in Fig. 2. The first  
peaks in $\epsilon_N$ for the models with 1.7 and 1.1 $M_{\odot}$ just before the 
ZAMS (zero age main sequence) are due to the reduction  in the original $^{12}$C content through the nuclear 
reactions $^{12}$C(p, $\gamma$)$^{13}$N($\beta^+$, $\nu$)$^{13}$C(p, $\gamma$)$^{14}$N. 
During the PMS stage the function is conserved for all models and also at the final stages
of WD, regardless of the initial mass of the model. Because the PMS duration depends on the
initial  masses, $\Gamma(M, \text{EOS})$ begins to increase for the most massive models first. 
For the least massive model the function is conserved almost throughout, except during 
a very short interval (log age $\approx$ 12.4), but the PMS value is recovered a little
later. 

  \begin{figure}
   \includegraphics[angle=-90,width=\columnwidth, totalheight=0.94\columnwidth]{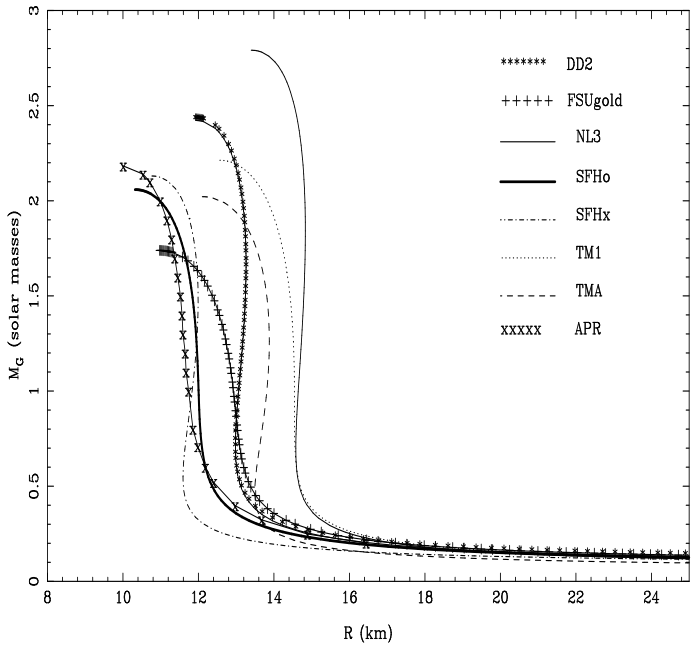}
   \caption{Mass-radius relation for neutron stars.  The adopted  EOS are
identified by the symbols displayed at the right upper corner of the figure. Only stable
models are shown.  }
   \end{figure}

By inspecting Fig. 2 we can deduce that there is a threshold below which 
$\Gamma(M, \text{EOS})$ is invariant. This limit is around  10$^{-3}$ ergs~s$^{-1}$~g$^{-1}$
(indicated by horizontal dashed line in the second panel of Fig. 2).  Above this
threshold, the stellar  mass distribution changes so drastically   due to the presence of
nuclear reactions and the subsequent chemical inhomogeneities that  $\Gamma(M, \text{EOS})$  
increases up to four orders of magnitude (not shown in the figure for the sake of clarity). 
 This occurs  mainly during some post-main sequence phases.  The behaviour of the 
configurations during these phases can be understood 
with the help of the virial theorem.  If the accelerative changes in the moment 
of  inertia are not large ($\frac{\partial^2I}{\partial t^2} \approx 0$), we have 
2 E - $\Omega$ = 0, where E is the total energy and $\Omega$ is  the generic 
gravitational potential energy given by $\alpha G M^2/R$. The form-factor 
$\alpha$ gives  a measure of mass concentration of a model, in a similar way as the
apsidal-motion constant does.  As the core contracts, $\alpha$  increases and  the 
external layers must expand to keep the gravitational potential energy constant. On the 
other hand, below the threshold the configurations are changed more smoothly and this  
function remains constant ($\approx$ 0.4) as in the cases of PMS, WD, NS,  and for the 
gaseous planets studied in the previous section.  Our criterion 
for quasi-static or stable evolution visible  in Fig. 2 
differs substantially from that used by Quiroga \& Claret (1992a) (see their figures 1, 3, and 4), who 
adopted  $0.38 \le \alpha\beta \le 0.44$ from the analytical approximations by Ferronsky 
et al. (1978). In addition, the  criterion by 
Quiroga \& Claret (1992a) does not relate the time evolution of $\alpha\beta$ with 
the nuclear reactions or with  the chemical inhomogeneities.

\section {Gravitational potential energy and the moment of inertia of
neutron stars: the role of the equation of state and the macroscopic stability
criterion}
The invariance of  $\left[\alpha\beta\right]_\mathrm{GR}/\Lambda^{0.9}(R)$ for
NS was shown in (Claret 2012) but using only  two EOS (Akmal,  Pandharipande  
\&  Ravenhall 1998,  Shen et al. 2011). Here, we extended these calculations to 
seven other EOS formulations (see subsection 2.3 and Fig. 3) to test whether this  
function is independent of the EOS  and also to refine the exponent of this function. 
First, we analysed each EOS separately and searched for the best exponent to the function 
$\Lambda(R)$ by means of the least-squares method. Next, we considered all EOS  as a whole 
and repeated the same procedure. The results are listed in Table 1 where the merit function
 $\chi^2/N$  is also tabulated,  with $N$ as the number of points. All resulting exponents 
are close to 0.80, the  strongest deviation comes from $FSUgold$, which predicts 0.86. 
However, this EOS also predicts an $M_{max}$ of 1.74 $M_{\odot}$, which contradicts  
with the value of the highest inferred  NS mass  (1.97  $M_{\odot}$) obtained by 
Demorest et al. (2010). Probably this EOS should be ruled out, but we decided to keep it in
our  calculation of the average exponent (see the last line of Table 1). 

The individual values of the exponent of   $\Lambda(R)$ as well as the 
average one obtained taking into account  all EOS together (Table 1) allow 
us  to consider that the function   $\Gamma(M, \text{EOS})$  is invariant 
for  NS and independent of the adopted EOS and of the stellar mass. 
 For WD and NS as the final products of stellar
evolution, we have  shown that they recover  the initial value of 
$\Gamma(M, \text{EOS})$ characteristic of  the PMS ($\approx$ 0.4), i. e., stars 
begin and end their lives in different ways but they recover  this fossil function 
at the last stages of stellar evolution. Whether this remains true for black holes has 
to be examined in future studies.

\begin{table}[t]
\caption[]{Neutron stars and EOS: exponents for the function $\Lambda(R)$ }
\begin{flushleft}
\begin{tabular}{lccc}
\hline
EOS           & $M_\mathrm{max}$ ( $M_\odot$)     & exponent & $\chi^2/N$   \\
\hline

$DD2$      & 2.4223&   0.79& 6.59E-6 \\
$FSUgold$  & 1.7392&   0.86& 2.13E-5\\
$NL3$      & 2.7911&   0.78& 2.76E-5\\
$SFHo$     & 2.0587&   0.82& 1.25E-5\\
$SFHx$     & 2.1301&   0.79& 2.66E-5\\
$TM1$      & 2.2130&   0.82& 4.18E-5\\
$TMA$      & 2.0217&   0.82& 6.52E-5\\
All EOS    & -    &    0.80& 4.31E-5\\
\hline 
\end{tabular}
\end{flushleft}
\end{table}

As we have seen, the function $\Gamma(M, \text{EOS})$ is  conserved for all NS  
models/EOS but $\left[\alpha\beta\right]_\mathrm{GR}$ is no longer constant and is a 
function of  mass (Fig. 4). Some interesting characteristics can be inferred from this 
figure. First, it can be noticed that for very low NS masses the effects of  
general relativity are not important and  $\left[\alpha\beta\right]_\mathrm{GR}$ tends 
to the Newtonian value. The second characteristic is related to the stability of NS 
models. All models displayed in Fig. 4 are in equilibrium, but this does not  
necessarily mean that they are stable. The main point concerning equilibrium configurations 
is that they may correspond to a maximum or minimum of energy with respect to 
compression/expansion.  For NS the necessary (but not sufficient) condition 
of stability is given by (e. g., Harrison et al. 1965)

  \begin{SCfigure*}
   \centering
   \includegraphics[angle=-90,totalheight=\hsize]{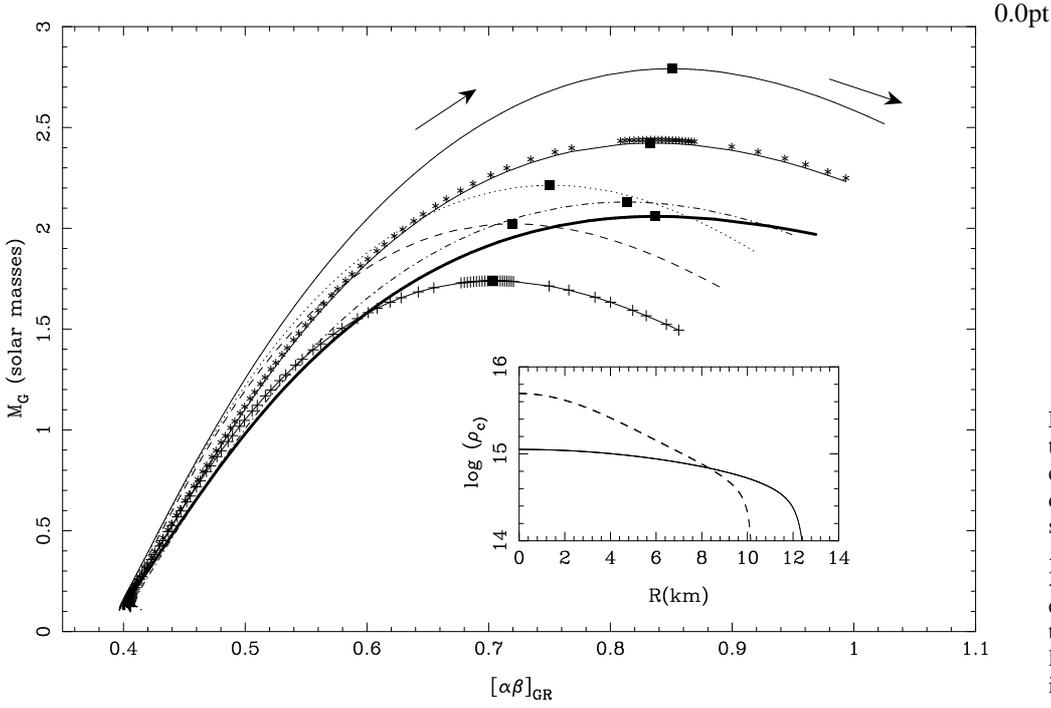}
\sidecaptionsep
   \caption{Mass-$\left[\alpha\beta\right]_{GR}$  relation  for 
neutron stars, including unstable models.  The two arrows indicate the direction 
of increasing central density and the full squares denote the $M_{max}$.  The same 
symbols as in Fig. 3. Lower right corner inset: mass-energy distribution for a stable  
(continuous line) and unstable (dashed line)  model with similar masses using  $DD2$ EOS.}
   \end{SCfigure*}
  \begin{figure}
   \includegraphics[angle=-90,width=\columnwidth, totalheight=0.94\columnwidth]{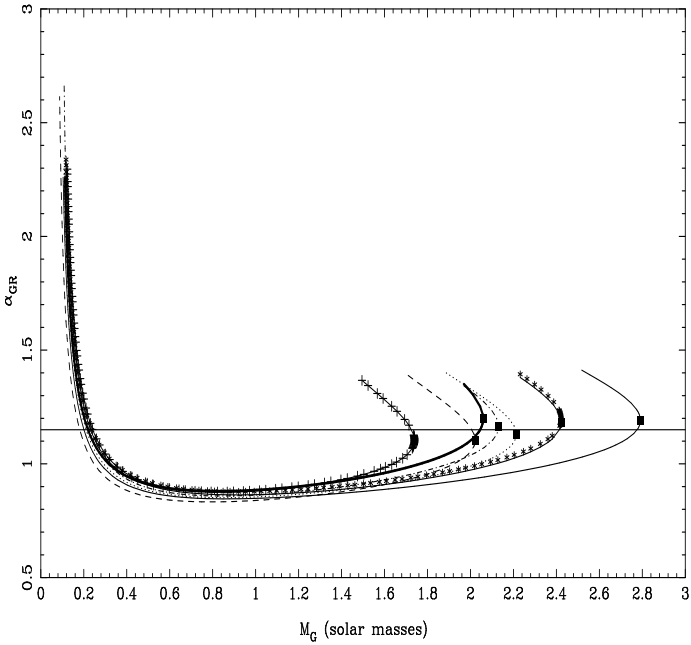}
   \caption{Mass-$\alpha_{GR}$  relation  for neutron stars. The same symbols as in 
Fig. 3. The horizontal full line indicates the average value of 
$\left[\alpha\right]_\mathrm{GRmax}$.}
   \end{figure}

\begin{equation}
 \frac{\partial M}{\partial \rho_c} > 0. 
\end{equation}

An equivalent criterion can be derived  considering the classical 
order-of-magnitude relations for the  pressure and density, $P \approx G
M^2/R^4$ and $\rho \approx M/R^3$, respectively. From these relations, we obtain  
(Cox \& Giuli 1968)

\begin{equation}
 M \approx \left(\frac{P}{G \rho^{4/3}}\right)^{3/2} \Rightarrow 
\frac23 \left(\frac{\partial {\rm ln} M} {\partial {\rm ln} \rho}\right)_S = 
\left(\frac{\partial {\rm ln} P}{\partial {\rm ln} \rho}\right)_S - \frac43
= 
\overline{\Gamma_1} - \frac43,
\end{equation}

\noindent
where the index $S$ refers to the entropy and $\overline{\Gamma_1}$ is an
average value over the  whole model. From Eq. 4 we derive the  condition  that a model 
will be stable for $\overline{\Gamma_1} > 4/3$. 

Now we use the Jacobi virial theorem to check the stability and derive the dimension of 
a NS without the need to use numerical calculations  like the integration of  
the TOV equations. We mainly follow the work by Fowler (1966). It can be shown that if 
we use the Jacobi stability criterion,  $\frac{\partial^2I}{\partial t^2} > 0$,   we
obtain the  inequality

\begin{equation}
\frac{R_S}{R_N} \left( 4 \zeta_1 ~~\overline {\Gamma_1 -1} + 2 \zeta_2
~~\overline 
{5 - 3\Gamma_1}\right)
< \left( \frac{3 ~~\overline {3\Gamma_1 -4}}{ 2 (5 - n)}\right),
\end{equation}

\noindent
where $R_S = 2 G M/c^2$ is the  Schwarzschild radius, $R_N$ is the radius of a 
neutron star, the factor $3/(5 - n)$ comes from the gravitational potential energy for a 
polytrope of index $n$, and $\zeta_n \approx  5.07/(5 - n)^2$. This function is connected 
with the post-Newtonian expansion of the proper internal energy and the gravitational 
potential energy of polytropes. Using the results of Eq. (4),   
$\overline {\Gamma_1} = 3/2$, $\zeta_1 = 0.31$,  $\zeta_2 = 0.55$, and $n$ = 2
we obtain $R_N > 4.7 R_S$. Introducing numerical values for 1  $M_{\odot}$, we obtain that 
a typical NS must have a radius of the order of 14 km, which  
agrees well with the average radius inferred from Fig. 3. Indeed,  for the more 
elaborated  calculations shown in Fig. 3, the radius  of a 1  $M_{\odot}$ NS 
model is between 12 and 15 km. 

Next,  we derive heuristically  an alternative macroscopic criterion of stability for 
NS based on the properties of their gravitational potential energies and  
moments of inertia through  $\left[\alpha\beta\right]_\mathrm{GR}$. We refer to Fig. 4, 
where  several NS models are displayed for seven EOS, including unstable
models. In this figure the arrows indicate the direction of increasing of $\rho_c$ and the
limiting masses are denoted by full squares. We first consider the configurations  to 
the left of the turning points. If a given configuration in this region  is compressed 
(higher $\rho_c$), the resulting configuration will present a  deficit of  gravitational 
mass if compared with its original position. The gravitational force will act to return 
it to the original position. Mutatis mutandis, if we decrease $\rho_c$, the star also 
returns to the initial position. For  models to the right of the turning points, this 
force will tend to move them from the original position.  Therefore, we 
introduce a macroscopic  criterion of stability:   the models will be stable only  
for positive slopes 

\begin{equation}
 { \frac{\partial M}{\partial [\alpha\beta]_{GR}}} > 0. 
\end{equation}

To illustrate this criterion of stability we plotted at the lower right corner of 
Fig. 4 the mass-energy distribution for two NS models with similar masses
(2.23  $M_{\odot}$) but located  to the left (stable) and  to the right (unstable) of the
turning point. The adopted EOS  was $DD2$.  The steeper profile near the centre 
corresponding to the model located  to the right of the turning point indicates that a 
perturbation in the star probably   induces it to collapse gravitationally to a 
black hole. 

As we have seen (Eq. 4), the critical value of $\overline{\Gamma_1}$ is $4/3$ for  
Newtonian configurations. However, the effects of the  general relativity on  
$\overline{\Gamma_{cri}}$ can be significant for NS, for example. Shapiro 
\& Teukolsky (1983) have shown that these  effects  can be expressed as

\begin{equation}
\overline{\Gamma_{cri}} = \frac43 + a\frac {GM}{{R c^2}},
\end{equation}

\noindent
where $a$ is a positive constant of the order of unity. Therefore, the  
effect of the General Relativity is to increase the value of $\overline{\Gamma_{cri}}$  
and  making NS more unstable.   On the other hand, we have shown in the beginning of this 
section that $\left[\alpha\beta\right]_\mathrm{GR}/\Lambda^{0.8}(R) \approx 0.4$ for 
all NS models. Combining Eq. 7 with the definition of the function  $\Lambda(R)$, we can  
show that 

\begin{equation}
\overline{\Gamma_{cri}} \approx \frac43 + a{\left(\frac{[\alpha\beta]_{GR} -
0.4}{[\alpha\beta]_{GR}}\right)}.
\end{equation}

\noindent
Equation 8 presents some interesting aspects. The first one is connected with
the Newtonian configurations. For PMS and WD sequences we have shown (Claret 2012) 
that $\left[\alpha\beta\right]_\mathrm{GR} \rightarrow 
\left[\alpha\beta\right]_\mathrm{Newt} \approx 0.4$.  Therefore, for these configurations 
the relativistic corrections are very small.  The other point   is that  which links  
Fig. 4,  Eq. 6, and Eq. 8. The behaviour of $\left[\alpha\beta\right]_\mathrm{GR}$  
for NS is  a key for understanding the stability of these stars and also for 
determining $M_{max}$. 

The general features shown in Fig. 4 as well as the alternative criterion of 
stability are due to the high compactness  of NS and are
directly related to  General Relativity.   As mentioned before, 
new observational data on NS masses gradually act as a discriminator and 
some EOS should be discarded. At our present level of knowledge, the EOS $FSUgold$ must 
be ruled out. In the framework of Fig. 4 the  limiting masses $M_{max}$  are 
located in a narrow range (0.72--0.85) of  $\left[\alpha\beta\right]_\mathrm{GR}$. In Fig. 5
we show the behaviour of  $\alpha_\mathrm{GR}$ as a function of mass. By inspecting this  
 figure we note that  $\alpha_\mathrm{GR}$  is  almost  independent of the EOS  at the 
turning points, i.e., for the $M_\mathrm{max}$. The value of $\alpha_\mathrm{GRmax}$ obtained 
with a least-square fitting is 1.15:  the strongest deviation  is 0.05. 
Surely, $\alpha_\mathrm{GRmax}$  can be refined with the contribution of improved EOS and 
new observational data of NS. 

In the present investigation, we  only included relativistic EOS models and, in  part  
the results for the non-relativistic APR EOS. The seven relativistic mean-field EOS of 
Hempel et al. cover  a broad range of possible neutron star mass-radius relations. 
Nevertheless, it would be good to include additional non-relativistic EOS and maybe also 
EOS with exotic degrees of freedom such as hyperons or quarks, to further validate the 
model-independency of the function $\Gamma(M, \text{EOS})$.

\section {Summary}

We have investigated some aspects of the stellar evolution concerning the  Jacobi 
virial equation, the function $\Gamma(M, \text{EOS})$, and the stability criterion. We summarise 
here the main results:

\noindent
1) We analysed the internal structure of gaseous planets with masses varying from 0.1 up 
to 50  $M_J$. We showed that the function $\Gamma(M, \text{EOS})$ is invariant for all models  
throughout the planetary evolution. In this way   we extended the invariance of the mentioned 
function to gaseous planets.  

\noindent
2) We found a connection between the strong variations of  $\Gamma(M, \text{EOS})$ during the 
intermediary evolutionary phases with the specific nuclear power. We also found a 
specific nuclear power threshold. Below this limit the  function is 
invariant ($\approx 0.4$) for PMS-WD models. 

\noindent
3) It was shown that the function $\Gamma(M, \text{EOS})$  for NS models is 
independent of the EOS and of the stellar mass.  

\noindent
4) A macroscopic stability criterion for NS models was introduced on the 
basis of the relativistic product $\left[\alpha\beta\right]_\mathrm{GR}$. 

\noindent
5) Although the observational data for NS does not allow us to 
distinguish among  all EOS yet, we determined that it seems to be a single  
maximum value of  $\left[\alpha\right]_\mathrm{GR}$ for the turning points. 

\noindent
6) We confirmed that regardless of the final products of the stellar evolution, NS  
 or  WD, they  recover the initial value of   
$\left[\alpha\beta\right]_\mathrm{GR}/\Lambda^{0.8}(R)  \approx 0.4$ acquired at
the PMS.  The case of black holes will be subject of a future study.

\begin{acknowledgements} 
We thank the anonymous referee for his/her useful comments and suggestions. 
We would like to thank B. Rufino and V. Costa for their comments. 
The Spanish MEC (AYA2009-10394, AYA2009-14000-C03-01) is gratefully  acknowledged 
for its support during the development of this work. M.~H. acknowledges support 
from the High Performance and High Productivity Computing (HP2C) project, and 
the Swiss National Science Foundation (SNF) under project number no. 200020-132816/1. 
M. H. is also grateful for support from ENSAR/THEXO and CompStar, a research networking 
program of the ESF.  This research has made use of the SIMBAD database, operated at the 
CDS, Strasbourg, France, and of NASA's Astrophysics Data System Abstract Service.
\end{acknowledgements}

\end{document}